\documentclass[a4paper]{article}

\usepackage{latexsym}

\newenvironment{proof}{ 
\begin{quotation}{\bf Proof:}\\*[\smallskipamount]}{\end{quotation}}
\newenvironment{remark}{ 
\begin{quotation}{\bf Remark:}\\*[\smallskipamount]}{\end
{quotation}}
\newenvironment{remarks}{
\begin{quotation}{\bf Remarks:}\\*[-\baselineskip]\noindent}{\end{quotation}}

\newenvironment{des}{\begin{enumerate} \begin{quotation}}
{\end{quotation} \end{enumerate}}

\newtheorem{ax}{A}
\newtheorem{df}{D}
\newtheorem{tre}{T}

\title{Axiomatic Foundations of Thermostatics}

\author {G. Puccini\thanks{Departamento de F\'\i{}sica, Universidad Nacional de
 La Plata, Argentina. Present address: Instituto de Neurociencias,
Universidad Miguel Hern\'andez, Apartado 18, 03550 San Juan de
Alicante, Spain.} , S. Perez Bergliaffa\thanks{
Centro Brasileiro de Pesquisas Fisicas (CBPF), 
Rua Dr. Xavier Sigaud 150, CEP 22290-180, Rio de
Janeiro, Brazil}   and H. Vucetich\thanks{
Observatorio Astron\'omico, Universidad Nacional de La Plata,
 Paseo del Bosque S./N., (1900) La Plata, Argentina}}

\begin{document}
\maketitle

\begin{abstract}
A realistic and objective
axiomatic formulation of Thermostatics for composite systems is
presented. The main feature of our axiomatics is that it is free of
empirical definitions. In particular,
the basic concepts of the theory, such as those of entropy, heat and
temperature, are characterized only by the axiomatic  basis and
the theorems derived from it. We also show that the concept of
(quasi)static process does not belong to the body of Thermostatics.
\end{abstract}

pacs: 05.70, 01.70.+w

\section{Introduction}

Throughout the history of science there have been many attempts to
formulate a consistent axiomatization of Thermodynamics. Although the
most widely known example is the work of Carath\'eodory
\cite{Caratheodory09}, the first attempt, by J. Gibbs, dates back to
1875 (For historical references, see Truesdell \cite{Truesdell84}).
Gibbs took at first energy, entropy, and absolute temperature as
primitive concepts, and later, in 1901, he tried to find a statistical
basis for the concept of entropy. Many axiomatic formulations of
Thermodynamics have been developed since then.  In most of the cases
the aim was to obtain a better understanding of the concept of
entropy. Some ``operational definition'' (formulated in terms of
measurable quantities) of entropy is at the heart of many of these
attempts.  For instance, Rastall \cite{Rastall70} emphazised that
``The advantage of our formalism is that it gives a more immediate
insight into the meaning of entropy and temperature.''  Another
example is Tisza \cite{Tisza61}, who put the theory of Gibbs in an
axiomatic format. Several authors
\cite{Landsberg56,Buchdahl60,Buchdahl62,BuchdahlGreve,Boyling72,Rastall70}
presented modern versions of Carath\'eodory's axiomatization%
\footnote{A critical analysis of the original formulation of
Carath\'eodory can be found in \cite{Truesdell84}.}.  These and many
other related works have (mistakenly, as we shall see) led to the
certainty that Thermodynamics rests on solid foundations.

 This approach to the axiomatization of Thermodynamics is based on the
derivation of the entropy function under certain assumptions. Some
authors \cite{Landsberg56,Buchdahl60,Buchdahl62,BuchdahlGreve,Boyling72,Rastall70} base 
their work on Caratheodory's
axiom of adiabatic inaccessibility, together with topological
postulates on the set of thermodynamic states.  Zeleznik \cite{Zelenik76}
postulates that Thermodynamics has an algebraic structure and
identifies the entropy as a purely mathematical concept.  The recent
work of Lieb and Yngvason \cite{Lieb99} is surely the most elaborated along
these lines.  They derive the entropy function
from certain axioms that satisfy the order relation of
adiabatic accessibility among equilibrium states.

Unfortunately, all these axiomatizations are pervaded by the
philosophical tenet known as operationism.  According to it, the
factual meaning of scientific constructs is specified only through
measurable quantities or processes that relate them.  A typical
example is that of temperature. We read in Ref. \cite{Prigogine67}
that ``$T$ is a universal function of the temperature of the system as
recorded by measuring some arbitrary property like electrical
resistance''. Many scientists are firmly convinced of the need of this
type of definition. For instance, Ref. \cite{Reidlich68} maintains that
``In physical science, a concept is defined by experimental
instruction. For quantitative concepts, the instruction must lead to a
Dedekind cut, {\em i.e.} to a measurement. It is necessary and
sufficient for the definition of a property $P$ that the prescribed
experimental procedures decides whether the value\ldots'' .  In spite
of this support, any axiomatics constructed on the operationist basis
inevitably leads to inconsistencies, because  operationalism fails
to realize the difference between the measurement of a magnitude (a
fact) and the definition of a concept (a purely conceptual process).
It is true that Thermodynamics, as any other phenomenological theory,
involves and interrelates observational variables. However, the
definition of these cannot be reduced to laboratory
operations. Moreover, as we shall see below several concepts with no
counterpart in the physical world play an important role in
Thermodynamics (for example, that of state)%
\footnote{A complete criticism of operationism can be found in
Ref. \cite{Bunge99}}.

In many of these formulations, guided by empiricist philosophies, heat
is considered as a concept that should not be defined because it
cannot be measured. Accordingly, Lieb and Yngvason \cite{Lieb99} state
that ``Another mysterious quantity is `heat'. No one has ever seen
heat, nor will it ever be seen, smelled or touched\ldots{} There is no
way to measure heat flux directly\ldots{} The reader will find no
mention of heat in our derivation of entropy, except as a mnemonic
guide.''  Another author says that ``It certainly would not do to say
that Thermodynamics deals with `heat'.  The latter has not been
defined, and indeed never be explicitly defined at all.''
\cite{Buchdahl60}.  However, Lavoisier and Laplace
\cite{Lavoisier1780} defined the concept of heat (in the framework of
``caloric'' theories), and devised accurate methods of measuring heat
fluxes. More remarkably, these methods served as the foundation basis
of Calorimetry, a branch of Thermodynamics that is essential to
research in many areas of science, such as physical chemistry or
low-temperature physics. We then take the view that any axiomatization
of Thermodynamics must try to define and clarify the concept of heat.

Another important remark is that the distinction between
Thermodynamics and Thermostatics is ``something not widely understood
even today'' \cite{Truesdell84}. However, we can assert that the main
interest of Thermostatics is the study of the equilibrium states of a
system. Consequently, the concept of {\em process} (in the sense of
time evolution of the state of a system) is alien to Thermostatics.
On the other hand, Thermodynamics deals with processes in which a
certain property, namely entropy, is produced in an irreversible way.
These processes cannot be spontaneously time reversed, and so time
plays an important albeit silent r\^ole in describing the evolution of
thermodynamical systems.

We think then that Thermodynamics and Thermostatics need a revision of
their foundations.  With this purpose in mind, we shall present here a
realistic and objective axiomatization of Thermostatics. We hope to report
more on the subject in a future paper on Thermodynamics.  Needless to say,
the axiomatic  format offers several advantages when compared to other
presentations. First, in it all the presuppositions of the theory are
made 
explicit from the beginning. This helps to avoid the intrusion of
elements  alien to physics ({\em e.g.,}  observers). Second, the
identification of the physical referents of the theory can be safely
performed. The referents are simply the arguments of the functions that
appear in the different statements \cite{Bunge74}. Third, the assignation
of meaning by means of semantical axioms precludes mistakes that originate
in  abuse of  analogy. Finally, the axiomatic  format
clarifies the structure of the theory and so it paves the way to the
deduction of theorems and the elimination of pseudotheorems.

In order to make contact with previous axiomatics, we must have some
criteria to discern between two axiomatic formulations of the same
theory.  A first noticeable difference originates in the freedom we
have on the election of the primitive basis. A second distinguishing
feature, which usually goes unnoticed in physics, are the
philosophical presuppositions adopted by each axiomatic system.  We
gave above some of the reasons why operationism, one of the most
common tenets among scientists, should be discarded from further
consideration in philosophy of science. We shall adopt instead a
realistic philosophy \cite{Bunge67,Bunge77,Bunge79}. We support
realism because, contrary to idealism, we assume that the entities we
shall study ({\em i.e.} the reference class of our axiomatization)
exist irrespectively of our sensory experience.  Finally, we shall
see that our axiomatization is objective, because no 
subject belongs to the domain of quantification of the bound variables
of the theory.

The structure of the paper is the following.  In the next section, we
give a brief summary of our ontologic presuppositions, which is based
on the realistic ontology of Bunge \cite{Bunge77,Bunge79}. In Section
\ref{Axiomatics} we present in detail the axiomatic basis of the
theory.  In Section \ref{Theorems} we show that several important
theorems can be derived from our basis. Finally, in the last section
we compare our presentation with previous ones, and we discuss our
results.

\section{Ontological Background}
\label{Ontology}

Originally, the interest in Thermodynamics arose with the need to
understand the transformation of energy by means of thermal
machines. A thermal machine is defined as a body composed of several
parts placed in a certain order and joined to each other, forming a
unit.  Contrary to what happens in simpler structures, if two parts
are changed in one of these machines, it loses some (or all) of its
features as a whole.  But the latter is no other that the ontological
concept of {\em system}. In particular, the formulation of
Thermostatics that we shall present here will refer to concrete
systems, that is to say, bodies connected to each other in some
way. Before giving a brief summary of the general theory of systems we
must characterize a few concepts.

    The basic concept of the realistic ontology developed in Refs.
\cite{Bunge77,Bunge79} is that of {\em substantial
individual}. Substantial individuals can associate to form new
substantial individuals, and they differ from the fictional entities
called {\em individuals without properties} precisely in that they
have a number of properties in addition to their capability of
association. Thus, concrete things are built from substantial
individuals $\xi$ together with their properties $P(\xi)$. In short,
denoting a thing by $x$, then we have: $x=\langle \xi, P(\xi) \rangle$
\footnote{The set of all things will be denoted by $\Theta $.}.
We will assume that any property $P$ of a thing $x$ is represented
by a mathematical function $F$, that is to say $F\stackrel{\wedge} {=}
P $.  We are interested in a formal characterization of a system. Some
definitions are needed first.

\begin{df}[Physical addition]
Let $x $ and $y$ be two different things. We can form a new thing $z$
by juxtaposing $x$ and $y$, that is to say: $z=x \dot{+} y$
\end{df}

\begin{df}[Action]
A thing $x$ acts on another thing $y$ if $x$ modifies the behavior of
$y$.  ($x \rhd y : x $ acts on $y$).  If the action is mutual it is
said that they interact ($x \Join y $).
\end{df}

\begin{df}[Connection]
Two things are connected (or bonded) if at least one of them acts on
the other.
\end{df}
In particular, the {\em bondage} of a set of things $A\subseteq\Theta$
is the set $B(A)$ of bonds (or links or connections) among them.

\begin{df}[Absolute Composition]
For every $x \in \Theta $, the composition of $x$ is:
$$
{\cal C}(x)= \{y \in \Theta / y \sqsubset x \},
$$
where ``$y \sqsubset x $'' designates ``$y$ is part of $x$''.
\end{df}

\begin{df}[Absolute environment]
\label{ambi}
 The environment of a thing $x$ is the set of things that are not
parts of $x$ but that are connected with some or all parts of $x$,
that is
$$
{\cal E}(x)=\{ y \in \Theta / y \not \in {\cal C}(x) \wedge (\exists
z)(z \in {\cal C}(x) \wedge (y \rhd z \vee z \rhd y))\}.
$$
\end{df}

\begin{df}[Absolute structure]
We will call structure of a thing $x$ the set of bonds ${\cal L}(x)$
among the components of $x$ and among $x$ and the things in its
environment.
\end{df}

A thing composed of at least two coupled components will be
called a system.  Formally,

\begin{df}[System]
A {\em system} $\sigma $ is a thing composed of at least two different
connected things.
\end{df}
We will adopt here a {\em minimal} model of system constituted by the
ordered triple:
$$
{\cal S}(\sigma)=\langle {\cal C},{\cal E},{\cal L} \rangle .
$$
Since in general we shall be dealing with systems composed of other
systems, we should keep in mind that a component can also be a
system itself. We need then to introduce the notion of {\em
subsystem}:

\begin{df}[Subsystem]
Let $\sigma$ be a system with $\langle {\cal C}(\sigma),{\cal E}
(\sigma),{\cal L}(\sigma)\rangle $.  Then a thing ${\sigma}_i$ is a
subsystem of $\sigma $ $({\sigma}_i \prec \sigma)$ if and only if:

\begin{enumerate}
\item ${\sigma}_i$ is a system, and
\item (${\cal C}({\sigma}_i) \subseteq {\cal C}(\sigma)) \wedge ({\cal
E} ({\sigma}_i) \supseteq {\cal E}(\sigma)) \wedge ({\cal
L}({\sigma}_i) \subseteq {\cal L}(\sigma)) $.
\end{enumerate}
\end{df}

In particular, Thermostatics is concerned with only one kind of
relationship, namely bond relationships. The bonds among subsystems
are called {\em internal bonds}, while the bonds between the system
and their environment are called {\em external bonds}.

Theoretical physics does not deal with concrete things but with
concepts, in particular with conceptual schemes called models.
For example, a certain quantity of oxygen is a thing, but in
Thermostatics we shall be concerned with a certain quantity of moles of
a given ideal gas, so the real gas is modeled by the ideal gas.

\begin{df}[Functional schema]
Let $\sigma_i $ be a system. A functional schema $b $ of $\sigma_i $
is a certain nonempty set $M$, together with a finite sequence of
mathematical functions $F_i $ on $M$, each one of which represents a
property of $\sigma_i $.  Shortly:
$$
b=\langle M,{\bf F} \rangle ,
$$
where
$$
{\bf F} = \langle F_i / F_i \mbox { is a function with domain } M \;
\wedge 1\leq i \leq p \rangle .
$$
\label{funsch}
\end{df}
Therefore, the thing $\sigma_i $ will be represented by the functional
scheme $b$, that is to say $b \stackrel {\wedge} {=} \sigma_i $.

\begin{remark}
    The auxiliary set $M$ generalizes in ontology the usual
    physical notion of {\em reference system}. It can be built as
    the conceivable state space (see below) of a reference thing
    $x_f$. In thermostatics, it can be constructed as a particular
    subset of the state space (see remarks to Ax.\ref{equi}).
\end{remark}

As we stated above, real things have properties. A detailed account of
the theory of properties is given in Ref. \cite{Bunge77}. We shall give here
only some useful definitions.
\begin{df}
$P\in {\cal P} \leftrightarrow(\exists \xi)(\xi \in S\wedge P\xi)$.
\end{df}
Here ${\cal P}$ is the set of all substantial properties, and $S$, the
set of substantial individuals. The set of
all the properties of a given individual is given by
\begin{df}
$P(\xi) = \{P\in {\cal P} / P\xi \}$.
\end{df}
We shall adopt the following classification of properties:
\begin{df}[Extensive and intensive properties]
\label{proext}
Let $P $ be a property of a composite system $\sigma = \sigma_1
\dot{+} \sigma_2 $, such that $F \stackrel{\wedge} {=} P $. $P$ is an
extensive property if and only if $F(\sigma_1 \dot{+} \sigma_2) =
F(\sigma_1) + F(\sigma_2) $.  Otherwise, the property $P $ will be
called intensive.
\end{df}

\begin{df}
Let $P $ be a property of $\sigma_i $ in an environment
$\overline{\sigma_i} $.  Then $\sigma_i $ is open with respect to $P $
if and only if $P $ is related to at least one property of things in
$\overline{\sigma_i} $. Otherwise $\sigma_i $ is closed regarding $P
$.
\end{df}

\begin{df}
A system is closed if and only if it is closed for every $P\in {\cal
P} $.
\end{df}

\begin{remarks}
\begin{enumerate}
\item Our definition of environment of a system ({\bf D\ref{ambi}}) as
the set of all the things coupled with the components of the system
requires a careful distinction of the different types of connections:
internal and external, chemical, mechanical, etc. These connections
must be taken from the background theories.
\item Notice that we talk about {\em relationships} among properties
of a thing and {\em connections} between things since the properties
are interdependent but not interacting.
\end{enumerate}
\end{remarks}

It is natural to assume that all things are in some state (without
specifying the type).  The state of a system can be characterized as
follows:

\begin{df}[State function]
\label{estado} 
Let $\sigma_i $ be a system modeled by a functional schema $b =
\langle M,{\bf F} \rangle $, such that each component of the function
$$ {\bf F}=\langle F_1,F_2,\ldots, F_p \rangle: M \rightarrow V_1
\times V_2 \times \cdots \times V_p $$ represents a property of
$\sigma_i $.  Then, each $F_i $ is a {\em state function} (or state
coordinate) of $\sigma_i $.  ${\bf F}$ is the {\em total state
function } of $\sigma_i $ and its value $$ {\bf F} (m)=\langle
F_1,F_2,\ldots, F_p \rangle(m)=\langle F_1(m),F_2(m),\ldots, F_p
(m)\rangle $$ for any $m \in M $it {\em represents the state} of
$\sigma_i $ in the representation $b$.
\end{df}

\begin{remarks}
\begin{enumerate}
\item A very important point concerning the notion of state is that
according to our definition {\bf D\ref{estado}} every state is a state
of some concrete thing. States are concepts that model the properties
of a physical system more or less accurately, but the converse is not
true: a physical system cannot be conceived as a bundle of properties
without physical support.
\item It must be noted that although the ontological concept of {\em state}
was defined in {\bf D\ref{estado}}, it is not the aim of Ontology to
specify a particular type of state.  More specifically, Thermostatics
only deals with {\em equilibrium states}, not with {\em quasiequilibrium
states}. We shall see below that these equilibrium states are
characterized by a particular value of the state function ${\bf F} $ of the
system.
\end{enumerate}
\end{remarks}

We shall call {\em conceivable state space} the set formed by the
Cartesian product of the range of each of the functions of ${\bf F}
$. This set will be denoted by $S(\sigma_i) $.

The states of the system are parametrized by the states of another
thing $x_f$ that qualifies as a reference frame, in the sense that $M
= S(x_f)$. That is, for each state $t \in M $ of the reference frame, the
system is in the state $s={\bf F}(t) $, where ${\bf F} $ is the state
function of the system.

Any restriction on the possible values of the components of ${\bf F} $
and any relationship among them is called a {\em law statement}. The set
of all the law statements involving $\sigma_i $ will be denoted
${\bf L}(\sigma_i) $.

\begin{df}[Lawful state space]
The subset of the range of ${\bf F}$ restricted by those law statement
in ${\bf L}(\sigma_i) $ will be called the {\em lawful state space} of
$\sigma_i $ in the representation $b$, and it will be denoted by
$S_{\bf L}(\sigma_i) $.
\end{df}

Every event is really a change of state of a system. If such a change of
state is carried out along a curve that characterizes the intermediate
states, the change we will called a \emph{transformation} and it will be
represented by a curve in the lawful state space of the system.

\begin{df}[Lawful transformations]
\label{tranf}
Let $S_L(\sigma_i) $ be the state space of system $\sigma_i $. Then
the family of {\em lawful transformations} of the lawful state space
into itself is the set $G_L$ of functions $g$ such that:
$$
G_L(\sigma_i)=\{ g:S_L(\sigma_i) \rightarrow S_L(\sigma_i) \wedge g
\mbox{ is compatible with the laws of }\;\sigma_i \}
$$
\end{df}
The transformation with starting point $\varphi $ and ending point
$\varphi'$ (with $\varphi $ and $\varphi'\in S_L(\sigma_i) $) will be
represented by the triple $\langle \varphi, \varphi ' , g \rangle $
where $ g \in G_L(\sigma_i) $ and $\varphi'=g(\varphi) $.

In the next section we will give an axiomatic presentation of
Thermostatics based on this ontological background. Supplementary
comments were added where we felt some
elucidation was needed.

\section{Axiomatics}
\label{Axiomatics}
\subsection{Formal Background}

\begin{enumerate}
\item Bivalent logic.
\item Formal semantics.
\item Mathematical analysis.
\end{enumerate}

\subsection{Material Background}
\label{material}

\begin{enumerate}
\item Macroscopic physics: classical mechanics, electromagnetism,
etc.
\item Chemistry.
\item General theory of systems.
\item Physical Geometry.
\end{enumerate}

\begin{remark}
We do not need to include Chronology ({\em i.e.} the set of theories of
time) in the background because time plays no role in Thermostatics.
\end{remark}

\subsection{Primitive Basis}
\label{Primitive}

The conceptual space of the theory is spanned by the basis {\bf B} of
primitive concepts, where $$ {\bf B}={\langle
\Sigma,\overline{\Sigma},{\cal F},\Phi,S,U,Q \rangle}. $$
The elements of
this basis will be partially characterized by the axiomatics of the theory
and the derived theorems. We shall class the axioms into three
different classes: mathematical ({\bf M}), physical ({\bf P}) and
semantical ({\bf S}), according to their status in the theory.

\subsection{Axioms}
\label{Axioms}

\begin{remark}
The geometrical notions used below (mainly that of volume) are taken
from the theory of physical geometry.
\end{remark}

\subsubsection*{Group I: Systems and states}

\begin{ax}[Thermostatical systems]
\begin{des}
\item {\bf [M]} $\Sigma $, $\overline \Sigma :\;$ nonempty sets.
\item {\bf [S]} $(\forall{\sigma} )_{\Sigma} \; (\sigma \stackrel{d}
{=} $ thermostatic system$). $%
\footnote{The symbol $\stackrel{d} {=}$ is used here for the relation
of denotation (see Ref. \cite{Bunge74} for details).}
\item {\bf [S]} $(\forall{\overline{\sigma}} )_{\overline{\Sigma}} \;
(\overline{\sigma} \stackrel{d} {=} $ environment of some thermostatic
system$). $
\end{des}
\end{ax}

\begin{ax}[Properties]
\label{properties}
\begin{des}
\item {\bf [M]} ${\cal F}: $ nonempty set.
\item {\bf [S]} $(\forall F)_{\cal F}\;(F \stackrel{\wedge}{=} $ property
of the system $).$
\end{des}
\end{ax}

\begin{ax}[States]
\label{estad}
\begin{des}
\item {\bf [M]} $\Phi $: nonempty set of functions.
\item {\bf [P]} $\forall{(\sigma, \overline{\sigma}) }_{\Sigma \times
\overline{\Sigma}}, (\exists \varphi)_{\Phi} \; / \; \varphi
\stackrel{\wedge}{=} {\bf F}(\sigma, \overline{\sigma}). $
\item {\bf [S]} $(\forall {\varphi} )_{\Phi} \; (\varphi \stackrel{d}
{=} $ equilibrium state of the system $). $
\end{des}
\end{ax}

\begin{remarks}
\begin{enumerate}
\item $\Sigma $ is the class of factual reference of Thermostatics. A
member $\sigma \in \Sigma $ represents an arbitrary thermostatic
system, for instance a body formed by other bodies
(subbodies or subsystems) in interaction. In next subsection we will
give a more accurate characterization of a thermostatic system.

\item Since to study a system it is necessary to conceptually
isolate it from the surroundings, the objects that are not part of
the system but interact with it in a noticeable way will be denoted
by $\overline \sigma $. They will be called environment of the given
thermostatic system and the set of all environments will be denoted by
$\overline \Sigma $.

\item {\bf A\ref{estad}} gives a mathematical characterization of each
$F \in \cal F $ as a function that depends neither on space
coordinates nor on time. Each $F$ depends solely on the system and its
environment, and has a subset of $\Re $ as range. These properties
need not be specified now. Since different systems usually require the
specification of different properties, this approach is valid for
different kinds of systems.

\item ${\cal F}$ should not be confused with {\bf F}. ${\cal F}$ denotes the
set of all functions representing properties of any thermostatic system
included in $\Sigma$. {\bf F} instead, denotes a list of all the functions
representing properties of a given thermostatic system.

\item From the axioms given above we can identify the set $M$ (see {\bf
D\ref{funsch}}) as a subspace of
$\Phi$. For instance $M = (p, V)$ or $M = (S, U)$ in the case of an
ideal gas. (Cf. {\bf A\ref{equi}} and {\bf D\ref{ThermoCoord},
D\ref{GenerCoord}.})

\item Note that the set $\Phi $ is the lawful state space of all
systems in $\Sigma $, while $S_L(\sigma) $ is the lawful state space
of a particular system $\sigma $.

\item We shall see that the main goal of Thermostatics is to predict
the properties of a given system in an equilibrium state reached by
means of some transformation $g $ which started from a different
equilibrium state. Moreover, it will be clear that a more detailed
description is obtained when the value of a certain functional
associated with each transformation can be established.

\item To avoid unnecessary complexity in the notation, the dependence of the
quantities on the reference thing $x_f$ (or the auxiliary set $M$)
will not be made explicit whenever possible.
\end{enumerate}
\end{remarks}

Next we shall characterize the basic unit of the composition of any
thermostatic system.  In most cases, a thermostatic system is a
composite system (commonly called heterogeneous system), and each of
these components is a homogeneous system, called phase or elementary
system.  The components of a homogeneous system, if any, are not 
thermostatic systems. This suggests the following definition:

\begin{df}[Elementary system]
Any thermostatic subsystem such that none of their components is a
system will be called elementary system (or phase). Formally:
$$
\sigma_i \stackrel{Df} {=} (\sigma_i)_\Sigma \wedge \forall x (x \in
{\cal C}(\sigma_i) \Rightarrow x \not \in \Sigma)
$$
with $\Sigma ^ * \subset \Sigma $,  the set of elementary systems.
\end{df}

\begin{ax}[Composition]
The composition of an elementary system is given by the set of
different chemical species present in it, that is to say:
$$
{\cal C}(\sigma_i)={\bigcup}_{i=1,\ldots ,r} e_i
$$
such that $e_i $ is a chemical species.  Thus, $\sigma_i $ will be
called elementary system or phase of $r$ components.
\end{ax}

\begin{remark}
    The concept of chemical species is taken from Chemistry,
    which is part of our material background.
\end{remark}

\begin{df}[Heterogeneous system]
Every composite system formed by two or more elementary systems
will be called heterogeneous.
\end{df}

Note that in the previous definition the composition of each phase has
not been mentioned.  This will allow us to define the concept of a 
simple system. We know that every system may be display a number of
phases, and that every system of $m$ phases can be transformed into
another system of $m'$ phases. Each of these phases may have the
same or different composition.  This motivates the following
definition:

\begin{df}[Simple system]
We call simple system $\sigma_i$ every elementary system and every
heterogeneous system formed by elementary systems having the same
composition. That is to say:
$$
\sigma_i=\dot{\sum_{j=1,\ldots,m}} \sigma_j \; , such\;\; that
\;\;\;{\cal C}(\sigma_j)= {\bigcup}_{l=1,\ldots ,r} e_l, \; \;
\;\;\;\; \forall j=1,\ldots, m.
$$
The system $\sigma_i$ is called a simple system of $m $ phases and $r $
components.
\end{df}

\begin{remark}
Note that every elementary system is a simple system, but all simple
systems are either elementary or a composition of them.
\end{remark}

The tacit assumption of Thermostatics is that each subsystem $\sigma_i
\in \Sigma^*$ is in equilibrium and has a number of properties that
do not depend on the space coordinates.  Accordingly, the formulation
is global, because the properties are constant in each $\sigma_i $.
Moreover, keeping in mind that each component of the state function
${\bf F}$ represents a property of $\sigma_i $, we define:

\begin{df}[Thermostatic coordinates]
\label{ThermoCoord}
Each of the state coordinates of the function ${\bf F} $ that represents
a property of system will be called a \emph{thermostatic coordinate}.
\end{df}

In accordance with the definition of extensive properties we can adopt
the following convention:
\begin{df}[Generalized coordinates]
\label{GenerCoord}
The functions $X $ that represent extensive properties will be called
{\em generalized coordinates}.  The set of all the generalized
coordinates will be denoted by {\bf X}.  Formally,
$$
{(\forall X)}_{\bf X} (\forall P)_{\cal P} ( X \stackrel{\wedge}{=}P
\wedge P \;{\mbox is\; extensive})\;,X \stackrel{Df}{=} \mbox{
generalized coordinate }
$$
\end{df}

\begin{df}[Thermostatic configuration space]
\label{confi}
For every $\sigma_i$, the space spanned by the generalized coordinates
is called the {\em thermostatic configuration space} and will be denoted
by $S(\sigma_i) $.
\end{df}

\begin{remarks}
\begin{enumerate}
\item It must be noted that all the quantities that specify a state
are time-independent.  In particular, the properties of elementary
systems are spatially and temporally constant. In particular, due to
the requirement of spatial constancy, inhomogeneous fields (such as
gravitational fields) are excluded from Thermostatics.

\item The only states that Thermostatics need are the equilibrium
states introduced in ({\bf A\ref{estad}}).  Consequently, properties
are defined only for such states.  This feature of Thermostatics has
been mistakenly extrapolated to Thermodynamics, where sometimes it is
considered that properties are not defined for states far from
equilibrium (For a discussion, see Ref. \cite{Truesdell84}). This
misconception, inherited from operationism, is caused by the belief
that properties are defined for the states in which it is possible to
measure them.  Moreover, we can draw no conclusions from Thermostatics
on this matter, because it deals only with equilibrium states.

\item {\bf D\ref{confi}} identifies the conceivable state space of a
system from the point of view of Thermostatics as the thermostatic
configuration space.

\item As a simple corollary of the above definitions, the conceivable
state space of a composite system $\sigma=\sigma_1 \dot{+} \sigma_2 $
is spanned by the Cartesian product $S(\sigma)=S(\sigma_1) \times
S(\sigma_2) $.

\item The definition of a closed system w.r.t. the property $P $
fixes the structure ({\em i.e.} the constraints) of the system. Let us
consider for example a system built up of $N $ subsystems, such that
each of these is open regarding $P_j \stackrel{\wedge} {=} X_j $, but
the composite system is closed regarding the same property. The
following condition must be fulfilled:
$$
\sum_j^N X_j=X^T = constant
$$
where $X^T $ is an invariant of the composite system for all possible
transformations.

\item Note that we have not defined ``reservoir", ``wall", or
``enclosure". Such concepts are only necessary in formulations based on
operationism.  For example, a usual definition of closed system is a
``system in an impermeable enclosure" \cite{Tisza61}.
\end{enumerate}
\end{remarks}

The state coordinates are not independent of each other. This is
clarified by the following definition:

\begin{df}[Thermostatic degrees of freedom]
We shall call thermostatic degrees of freedom ($\cal N $) the minimum
number of independent state coordinates $F_i $ that should be
specified to determine univocally the equilibrium state of the system.
\end{df}

We have characterized the systems and classified their properties in
the equilibrium states.  We consider next two other physical
properties of the primitive base:

\subsubsection*{Group II: Fundamental properties}

\begin{ax}[Entropy]
\label{entro}
\begin{des}
\item {\bf [M]} $\forall{(\sigma_i , \overline \sigma_i)}_{\Sigma
\times \overline {\Sigma}}\; (\exists S)_{\cal F} (S:\Sigma \times
\overline{\Sigma} \rightarrow \Re^{+}).$
\item {\bf [P]} $S \in \bf X$.
\item {\bf [S]} $S(\sigma_i,\overline{\sigma_i}) \stackrel{\wedge} {=}
$ entropy of the system $\sigma_i $ when it interacts with the
environment $\overline {\sigma_i} $.  Notation: $X_0\stackrel{d}{=}S
$.
\end{des}
\end{ax}

\begin{ax}[Energy]
\label{energ}
\begin{des}
\item {\bf [M]} $\forall{(\sigma_i , \overline {\sigma_i})}_{\Sigma
\times \overline {\Sigma}}\; (\exists U)_{\cal F} (U:\Sigma \times
\overline{\Sigma} \rightarrow \Re^{+}).$
\item {\bf [M]} $U $ is a continuous, differentiable and monotonically
growing function \mbox{of $S$.}
\item {\bf [P]} $U \in {\bf X}$.
\item {\bf [S]} $U(\sigma_i,\overline{\sigma_i}) \stackrel{\wedge} {=}
$ internal energy of the system $\sigma_i $ in the presence of
$\overline{\sigma_i} $.
\end{des}
\end{ax}

\begin{remark}
Although these axioms refer to states and properties of simple systems
$\sigma_i $, they can be extended to the more general case of composite
systems. This extension is guaranteed by {\bf A\ref{entro}} and {\bf
A\ref{energ}}, which assert that entropy and energy are extensive
properties.
\end{remark}

The equilibrium states are represented by the state function ${\bf F}
$. We now proceed to list the axioms that relate the independent
properties of the system to those states. That is to say, we list the
axioms that characterize both mathematically and physically the
equilibrium states.

\subsubsection*{Group III: Equilibrium states}

\begin{ax}[Equilibrium representation]
\label{equi}
\begin{des}
\item {\bf [M]} The equilibrium states of every thermostatic system
are completely specified by the generalized coordinates. This means
that the state function is given by
$$
{\bf F}=\langle X_0,\ldots,X_n,X_{n+1},\ldots,X_p \rangle ,
$$
where the first $n+1$ coordinates are independent and they determine
the state of the system.  The remaining $p-n$ coordinates are
dependent on the ones in the first group.
\end{des}
\end{ax}

\begin{ax}[``Second Law'']
\label{condi}
\begin{des}
\item {\bf [P]} For every closed composite system $\sigma $, the
values taken by the
independent generalized coordinates
in the equilibrium state $\varphi ' $  are those that minimize the
function $U$ of the system, taking into account the values in the
state $\varphi $ with internal bonds.
\end{des}
\end{ax}

\begin{ax}[``Third Law'']
\label{nerst}
\begin{des}
\item {\bf [P]} The entropy of any thermostatic system is null in the state
for which ${(\frac{\partial U} {\partial X_0}) }_{X_1,\ldots,X_n}=0 $.
\end{des}
\end{ax}

\begin{remarks}
\begin{enumerate}
\item {\bf A\ref{equi}} is of fundamental importance because it states
that equilibrium states are independent of each other. In other words,
Thermostatics studies only those systems in which the inherited
changes are not contemplated. Also, it restricts the configuration
space $S(\sigma) $ since it assures the interrelation among dependent
and independent coordinates.

\item The set of $n+1$ independent coordinates in  {\bf A\ref{equi}}
spans the auxiliary set $M$ and introduces an implicit selection of
the reference thing $x_f$.

\item In this representation, the internal energy is a dependent
generalized coordinate. This means that it can be written as
$U=U(X_0,\ldots,X_n) $. Then {\bf A\ref{condi}} settles the restriction
that in any equilibrium state the form $dU $ must be
zero. That is to say,
$$
\left(\frac{\partial U}{\partial X_j}\right) =0 \;\;\; \; j=0, \ldots, n.
$$
We shall see below that the  minimum condition leads
to a classification of the equilibrium states regarding its stability.

\item {\bf A\ref{condi}} and {\bf A\ref{nerst}} are usually called
second and third principle of Thermodynamics, respectively.

\item {\bf A\ref{condi}} refers to closed composite systems because
Thermostatics is concerned with the determination of the final state
reached after reducing the number of bond relationships among
subsystems (internal bonds).

\item Thermostatics deals only with initial and final equilibrium
states: there is no room in the theory for real processes because any
process is nothing but a sequence of temporally ordered lawful states.
Therefore, the kind of process that the theory deals with are
fictitious processes, if only because it takes no time for them to
happen.  Let us remark that quasiequilibrium states or states that
``differ infinitesimally'' from equilibrium states are alien to
Thermostatics.  This eliminates from the theory the
quasistatic processes, present in almost every elementary
introduction to Thermostatics.

\item It follows from the previous remark that the concept of
``irreversibility'', associated to real processes does not play any
role in Thermostatics.  Also, the notion of a ``spontaneous process''
does not have any semantic content in the theory because time is
absent from it.
\end{enumerate}
\end{remarks}

Next we give some more definitions that will be useful in the
following.

\begin{df}[Fundamental equation]
\label{ecfun}
For every simple system $\sigma_i $, we will call fundamental equation
in the energy representation the expression of the internal energy in
terms of the remaining generalized coordinates of state, that is to
say: $U=U(X_0,\ldots,X_n) $
\end{df}

\begin{df}[Generalized forces]
\label{fuer}
For every simple system $\sigma_i $, we define the generalized force
$Y_j$ as
$$
Y_j \stackrel{Df} {=} {\left(\frac{\partial U}{\partial {X_j}}\right)
}_{X_0,\ldots,X_n}.
$$
In particular, $Y_0 $ will be denoted by $T$ and will be called {\em
temperature}:
$$
T \stackrel{Df}{=} {\left(\frac{\partial U}{\partial S}\right)}.
$$
\end{df}

\begin{remark}
Let us point out that temperature here is only a conventional name.
We need a semantic postulate (or rule of correspondence) in order to
confer some physical meaning to it.
\end{remark}

\begin{df}[Equation of state]
Every expression that gives the generalized forces in terms of the
independent generalized coordinates $Y_i=Y_i(X_0,\ldots,X_n) $ will be
called a {\em  state equation of the system}.
\end{df}

\begin{df}[Generalized work]
\label{traba}
The {\em generalized work} $W_j $ associated with the transformation
$g $ on $S_L(\sigma_i)$ is given by the following integral:
$$
W_j=\int_g Y_j dX_j.
$$
\end{df}

\begin{remarks}
\begin{enumerate}

\item It is not the goal of Thermostatics to give the functional form
of $U $. Therefore, the fundamental equation must be conjectured or
perhaps taken from some other branch of Physics. Nevertheless, once
its expression has been obtained, the formalism provides the means to
obtain results that can be contrasted empirically.

\item The assumption that $U $ would be a monotonically growing
function of $S$ can be written locally as $\frac{\partial U}{\partial
S} > 0 $, and globally as follows:
$$
U(X_0'',\ldots,X_n) \geq U(X_0',\ldots,X_n)\;,\;\; X_0'' \geq X_0' .
$$

\item Neither empirical temperature nor empirical entropy need to be
defined in this axiomatization.

\item Depending on the type of system and on the choice of the
representation, volume ($V $), longitude ($L $), surface ($\Gamma $),
magnetic moment ($M $), electric charge ($q $), and number of moles ($N $)
will be generalized coordinates.  The corresponding generalized forces
$Y_i $ will be pressure ($P $), traction ($\tau $), superficial
tension ($\gamma $), intensity of magnetic field ($B $), electromotive
force ($\varepsilon $), and chemical potential ($\mu $). All these
concepts are taken from theories in our background.

\item There are as many generalized forces as generalized
coordinates. In other words, there are as many state equations as
generalized state coordinates. For example, in the case of pure
substances, we have the ``pressure'' and the ``caloric'' equations of
state \cite{Callen85}. Not all of them, however, are independent.
\end{enumerate}
\end{remarks}

We have defined the functional associated with the transformations $g
$ (see {\bf D\ref{traba}}).   To complete our axiomatics we
must characterize an important functional in Thermostatics (which
represents the energy transfer among subsystems) and its relationship with
the previously defined functionals:

\subsubsection*{Group IV: Energy transfer}

\begin{ax}[Temperature (``Zeroth Law'')]
\label{temp}
\begin{des}
\item {\bf [S]} $T \stackrel{\wedge} {=} $ system temperature.
\end{des}
\end{ax}

\begin{ax}[Heat]
\label{calor}
\begin{des}
\item {\bf [M]} $\forall{(\sigma_i, \overline {\sigma_i}) }_{\Sigma
\times \overline {\Sigma}} \; (\forall g)_{G_L(\sigma_i)} (\exists
Q)(Q:\Sigma \times \overline \Sigma \times G_L \rightarrow \Re ). $
\item {\bf [S]} $Q(\sigma_i, \overline \sigma_i, g)\stackrel{\wedge}
{=} $ heat exchanged between $\sigma_i $ and $\overline \sigma_i $,
associated with the possible transformation represented by the function $g
$. Notation: $W_0 \stackrel{d}{=}Q$.
\end{des}
\end{ax}

\begin{ax}[``First law'']
\label{cons}
\begin{des}
\item {\bf [P]} $\forall{(\sigma_i, \overline {\sigma_i}) }_{\Sigma
\times \overline {\Sigma}} \; (\forall g)_{G_L(\sigma_i)},  \Delta
U=Q + \sum_j W_j=\sum_{j=0\ldots n}W_j $.
\end{des}
\end{ax}

\begin{remarks}
\begin{enumerate}
\item {\bf A\ref{calor}} characterizes heat as a form of energy
transfer among the system and its environment, which depends on the
transformation $g$. In other words, heat is interpreted just as
another form of interaction between systems.

\item Note that there is no mention to measurements of any kind in the
definition of heat.

\item {\bf A\ref{cons}} represents the conservation of energy (usually
called the first principle).  It must be realized that this is a
fundamental natural law, and consequently cannot be taken as a
definition of heat, as is often stated
\cite{Reidlich68,Rastall70,Boyling72,Callen85}.
\end{enumerate}
\end{remarks}

\section{Theorems}
\label{Theorems}

In this section we will give some theorems that can be deduced
from our axiomatic  basis.

\begin{df}[$n$th order homogeneous function]
\label{homo}
A function $\phi $ of the set of independent variables $(X_1, \ldots,
X_j) $ is homogeneous of $n$th order if it satisfies the following
condition:
$$
\phi(\lambda X_1,\ldots,\lambda X_j)=\lambda^{n}\phi (X_1,\ldots,X_j).
$$
where $\lambda$ is an arbitrary real number.
\end{df}

\begin{tre}[Extensivity criterion]
Every extensive property is a homogeneous function of first order,
w.r.t the independent generalized coordinates,
$$
X(Y_i,\ldots,Y_l,\lambda X_{l+1},\ldots,\lambda X_n)=\lambda
X(Y_i,\ldots,Y_l, X_{l+1},\ldots, X_n).
$$
\end{tre}

\begin{proof}
From {\bf D\ref{proext}} and {\bf D\ref{homo}}.
\end{proof}

\begin{tre}[Intensivity criterion]
Every intensive property is a homogeneous function of order zero
w.r.t. the independent generalized coordinates.
$$
Y(Y_i,\ldots,Y_l,\lambda X_{l+1},\ldots,\lambda X_n)=Y(Y_i,\ldots,Y_l,
X_{l+1},\ldots, X_n)
$$
\end{tre}

\begin{proof}
From {\bf D\ref{proext}} and {\bf D\ref{homo}}.
\end{proof}

\begin{tre}[Euler]
\label{euler}
Every generalized coordinate $X $ can be written as
$$
X=\sum_{i=1,\ldots,n} {X_i} {\left(\frac {\partial X}{\partial
{X_i}}\right)}_{{X_j}},\;\;\;j\neq i .
$$
\end{tre}

\begin{tre}[Euler theorem applied to the energy]
$$
U=\sum_{i=0,\ldots,n} X_i Y_i .
$$
\end{tre}

\begin{proof}
Set $X=U $ in Euler's theorem, {\bf T\ref{euler}}.
\end{proof}

\begin{tre}[Positivity of $T$]
$\forall (\sigma_i, \bar\sigma_i)_{\Sigma\times\bar\Sigma}(\exists T)(T
\geq 0). $
\end{tre}

\begin{proof}
It follows from the fact that $U $ is an increasing
function of $S $.
\end{proof}

\begin{tre}[Subaditivity of $U$]
The function $U $ is subadditive. That is to say, for a composite
system $\sigma=\sigma_1 \dot{+} \sigma_2 $,
$$
U(\sigma,\overline \sigma) \leq U(\sigma_1,\overline \sigma_1) +
U(\sigma_2,\overline \sigma_2),
$$
where $\overline \sigma $ denotes the environment of the composite system
in the new equilibrium state without bonds.
\end{tre}

\begin{proof}
It follows from {\bf A\ref{condi}}.
\end{proof}

\begin{tre}[Convexity of $U$]
\label{conve}
$U$ is a globally convex function; {\em i.e.} for any changes $\Delta
X_j $:
$$
2U(X_0,\dots,X_n) \leq U(X_0-\Delta X_0,\dots,X_n-\Delta X_n)+U(X_0
+\Delta X_0,\dots,X_n +\Delta X_n).
$$
\end{tre}

\begin{proof}
It follows as a consequence of the subadditivity and extensivity of $U$
\cite{Galgani68}.
\end{proof}

\begin{remark}
The mathematical properties of energy expressed by the above theorems
play a fundamental r\^ole in Caratheodory-like axiomatizations.
\end{remark}

\begin{tre}[Gibbs-Duhem]
The variables associated to every system with fundamental equation
$U=U(X_0,\ldots,X_n) $, satisfy the following condition:
$$\sum_{i=0,\ldots,n} X_i dY_i=0$$
\end{tre}

\begin{proof}
Compare the differential of the fundamental equation {\bf
D\ref{ecfun}} with Euler's theorem applied to the energy.
\end{proof}

The following theorem settles the number of thermostatic degrees of
freedom for a system composed by open subsystems with
respect to matter exchange and not subjected to electric, magnetic or
gravitational fields.

\begin{tre}[Phase rule]
The number of thermostatic degrees of freedom of a system $\sigma $
composed of $m$ elementary systems $\sigma_i $ in equilibrium and $r$
components is ${\cal N}=r-m+2 $.
\end{tre}

\begin{proof}
    From the definitions of phase, component and thermostatic
    coordinates.
\end{proof}

\begin{remark}
Obviously in the general case of systems composed by subsystems that
are open with respect to other properties, $\cal N $ will be given by
the nature of the system and by the number of bond relationships. That
is to say, if $n_v$ thermostatic coordinates are subjected to $n_l $
bonds relationships, $\cal N $ will be given by ${\cal N}=n_v - n_l$.
\end{remark}

Since the function $U $ of a composite system can have several minima, we
can give a classification of the different equilibrium states keeping in
mind that the condition of minimum of {\bf A\ref{condi}} can be local or
global:

\begin{tre}[Equilibrium criterion]
 Consider a closed composite system with fundamental equation
$U=U(X_0,\ldots,X_n) $.  The  equilibrium
state $\varphi '$ is a local minimum if $d^2U > 0$ at $\varphi '$.
This is satisfied
if and only if the Hessians at $\varphi '$ are positive, that is to say:
$$
H_j > 0 \;\;{\mbox  for }\;\;j=0,\ldots,n.
$$
\end{tre}

\begin{tre}[Global equilibrium criterion]
\label{glo}
 Consider a closed composite system with fundamental equation
$U=U(X_0,\ldots,X_n) $.  The
equilibrium state $\varphi ' $ is a
global minimum or not according to the convexity of the
function $U $ at $\varphi '$.
\end{tre}

\begin{proof}
Follows from {\bf A\ref{condi}}.
\end{proof}

\begin{df}[Classification of the equilibrium states]\hfill

\begin{enumerate}
\item An equilibrium state $\varphi ' $ will be called {\em stable}
iff $dU=0 $, $d^2U > 0$ and {\bf T\ref{glo}} is satisfied.
\item An equilibrium state $\varphi ' $ will called {\em metastable}
if $dU=0 $, $d^2U > 0 $ and if {\bf T\ref{glo}} is not satisfied.
\item A equilibrium state $\varphi ' $ will be called {\em critical}
if $dU=0 $, $d^2U=0 $ and it satisfies {\bf T\ref{glo}}.
\item A equilibrium state $\varphi ' $ will be called {\em unstable}
if $dU=0 $, $d^2U < 0 $.
\end{enumerate}
\end{df}

\begin{remark}
In the previous classification, the first three cases belong to
$S_L(\sigma) $, while the unstable equilibrium states are beyond the scope
of Thermostatics.
\end{remark}

Up to now we have been working with a functional model $b=\langle
M,{\bf F}\rangle $ where the components of ${\bf F} $ were all
generalized state coordinates.  That is to say, whenever we use the
fundamental equation $U $, we work in the so called {\em energy
representation}.  It is possible to rewrite the results in any other
representation that uses generalized coordinates and generalized
forces (or only generalized forces) by means of a change of
representation.

\begin{tre}[Change of representation]
Given a simple system $\sigma_i $ with fundamental equation
$U(X_0,\ldots,X_n) $, it is possible to represent the system
by means of a function $R(X_0,\ldots,X_l,$ $Y_{l+1},\ldots,Y_n) $ in
order to obtain an equivalent representation of the original one, by
means of the following Legendre transformation:

$$ U=\langle M,{\bf F}\rangle \rightarrow f(U)=\langle M',f({\bf
F})\rangle , $$ $$ U=U(X_0,\ldots,X_n) \rightarrow
f(U)=R(X_0,\ldots,X_l,Y_{l+1},\ldots,Y_n)=U-\sum_{j=l+1,\ldots,n}X_j Y_j ,
$$ such that $$ {\left(\frac{\partial R}{\partial {{X_i}}}\right)}=Y_i
\;\;, \;\;i=0,\ldots,l , $$ $$ {\left(\frac{\partial R}{\partial
{{Y_i}}}\right)}=-X_i \;\;,\;\;i=l+1,\ldots,n.$$
\end{tre}

\begin{tre}[Equilibrium condition in an arbitrary representation]
\label{gen-equil}
 The potential $R $ is
a convex function of the generalized coordinates and a concave
function of the generalized forces.
\end{tre}

\begin{proof}
See Ref. \cite{Galgani69}.
\end{proof}

\begin{remarks}
\begin{enumerate}
\item The function $R $ in the new state coordinates will be called
\emph{thermodynamic potential}.

\item This theorem warrants that under any change of representation,
the equilibrium states will be completely specified by $l+1$
generalized coordinates and by $n-l $ generalized forces.

\item The $n $ generalized forces conjugated to the $n $ generalized
coordinates form a total of $n $ couples $(X_i,Y_i) $ whose product
has energy dimensions. All the possible combinations of state
coordinates, taking one of each conjugated couple, is equal to $2^n $.
That is to say, in a system with $n $ coordinates we can choose among
$2^n $ thermodynamic potentials.

\item {\bf T\ref{gen-equil}} extends the condition of stability to
other representations.
\end{enumerate}
\end{remarks}

\begin{df}[Phase transition]
Let $\sigma$ be a system with thermodynamic potential
$$
R(X_0,\ldots,X_l,Y_{l+1},\ldots,Y_n) .
$$
Whenever some discontinuity
exists in one or more first derivatives of $R $, we say that $\sigma$
undergoes a phase transition.
\end{df}

\begin{df}[Continuous phase transition]
Let $\sigma$ be a system with thermodynamic potential
$R(X_0, \ldots, X_l, Y_{l+1}, \ldots, Y_n) $.  Whenever the first
derivatives are continuous but the second derivatives of $R $ are
discontinuous or infinite, we say that $\sigma$ undergoes a continuous
phase transition.
\end{df}

\begin{tre}[Maxwell relations]
\label{MaxRel}
For every system with thermodynamic potential
$$
R(X_0, \ldots, X_l, Y_{l+1}, \ldots, Y_n),
$$
the following conditions
are satisfied:
$$
{\frac{\partial X_j}{\partial {{Y_k}}}}={\frac{\partial X_k}{\partial
{{Y_j}}}} \;\;\; j,k \leq l,
$$
$$
{\frac{\partial X_j}{\partial {{X_k}}}}=-{\frac{\partial Y_k}{\partial
{{Y_j}}}} \;\;\; j \leq l \; k > l,
$$
$$
{\frac{\partial Y_j}{\partial {{X_k}}}}={\frac{\partial Y_k}{\partial
{{X_j}}}} \;\;\; j,k > l,
$$
where all the variables $X_0,\ldots,X_l,Y_{l+1},\ldots,Y_n $ remain
constant except the variable which is being differentiated.
\end{tre}

\begin{proof}
From the equality of the mixed second derivatives of the thermodynamic
potential $R $.
\end{proof}

\begin{remark}
    Maxwell's relations ({\bf T\ref{MaxRel}}) define a symplectic
    structure on the state space of a thermostatic system, whose
    fundamental form is \cite{Chen99}:
    \begin{displaymath}
    \theta = d U - \sum_j Y_j dX_j = \delta Q
    \end{displaymath}
    the last member being a usual notation for ``infinitesimal heat
    transfer''. We see then that this geometric structure, which is
    basic in Carath\'eodory-like formulations, is secondary
    in our axiomatics.
\end{remark}

\section{Discussion}
\label{Discussion}

We have given an axiomatic presentation of Thermostatics that is
realistic and objective. One of the effects of the assumption of a
realistic ontology is the complete absence of definitions that
(inconsistently) appeal to measurements in our axiomatics.  Also, many
of the basic concepts that are presupposed by other axiomatizations
have been specified here ({\em e.g.} those of system, subsystem,
environment, state, etc).

The conceptual space of the theory is generated by a 7-tuple in which
the set of systems is the unique factual reference.  The other
primitive elements are concepts of two types. The first type are
properties (or groups of properties) of systems. The second type are
those associated with the interaction of systems with the environment.
These fundamental properties that are part of the primitive basis are
entropy, internal energy and heat (introduced here as an
interaction). They allow us to define Thermostatics (and maybe
Thermodynamics too)
as the science of heat and temperature, which is in this
axiomatization a defined concept.  Equilibrium states have also been
introduced as primitives, since Thermostatics deals with this sole
class of states.  This entails that properties are necessarily defined
for this states, but it does not mean that they cannot be extended for
non-equilibrium states, such as those studied by Thermodynamics.

It is also clear from our axioms that Thermostatics deals neither with
processes nor with generalizations of them (like the quasistatic
processes, originally defined as {\em infinitely small changes of
system carried out at infinitely large time, to allow the system to be
in equilibrium in all instants of time}).

Most of present-day axiomatizations incorporate concepts that have no
room in a realistic philosophy of science. Among others, we can cite
those of ``empirical temperature'' , ``empirical entropy'', ``partitions''
and ``enclosures'' (used to characterize the connection among
subsystems).  Note that we have not introduce neither empirical
definitions nor experimental devices to justify them. However, these
should be consistent with the definitions given here.

The concept of heat was introduced here in accordance with historical
tradition \cite{Lavoisier1780} and with the definitions implicit in
the usual methods of calorimetry. The introduction of definitions and
axioms in the present formulation has been guided by empirical
knowledge, but they have no logical or epistemological dependence on
it. The reverse is true: the design of experiment (and of
technological devices) is guided by the theories (or, at least,
theoretical hypotheses) that are being tested (or applied).

Let us remark that formalizations based on Caratheodory's approach do
not offer an improvement on the understanding of the entropy
concept. The assignment of {\em properties} to the state space does
not provide a deepening of the concepts, but (perhaps) a deepening
into the mathematical foundations.  To achieve understanding (which is
a psychological category) it is necessary to give an explanation
(epistemological category) that involves some physical mechanism of
interaction (ontological category). Because of its very nature, no
mechanism is assumed in Thermodynamics, except that of heat exchange.
Indeed, Thermostatics is a very versatile theory because it is an
example of a {\em black box theory}. That is to say, no hypothesis is
made about the internal mechanisms acting in the system under study.
The relationship among observational variables is structural since the
interaction mechanism is at the level of systems, and therefore we can
ignore the particular internal details. These are left to deeper
theories like statistical mechanics. In this way, depth is achieved by
conjecturing some interaction mechanism and testing its consequences.

However, given its extreme generality, the theory cannot be
put to the experimental test. (The same holds for any other
hypergeneral theory, such as Hamiltonian dynamics and general field
theory.) To test the theory we must enrich it with a set of constutive
equations (such as the ideal gas law) that specify the nature of the
``stuff'' (or matter) that the system of interest is made of. But of
course, such enrichment (or specialization) goes beyond the
foundations of physics, which is only concerned with the entire
genera of physical things.

\section{Acknowledgements}

The authors would like to thank G. E. Romero and P. Sisterna for
helpful comments and M. A. Bunge for a critical reading of the
manuscript and important advice. HV is member of CONICET and
acknowledges support of the University of La Plata. SEPB acknowledges
support from CLAF and CONICET. GP acknowledges support from FOMEC
scholarship program.


\begin{thebibliography}{00}
\bibitem{Caratheodory09} {Carath\'eodory~C.} {\em Math. Ann.} {\bf 67}
(1909) 355. 
\bibitem{Truesdell84} {Truesdell~C.} {\em Rational
Thermodynamics}, 2nd.Edn. (Springer-Ver\-lag, New York) (1984).
\bibitem{Rastall70} {Rastall~P.} {\em J. Math. Phys.} {\bf 11} (1970) 2955.
\bibitem{Tisza61} {Tisza~L.} {\em Ann. Phys.} {\bf 13} (1961) 1.
\bibitem{Landsberg56} {Landsberg~P.} {\em Rev. Mod. Phys.} {\bf 28} (1956) 363. 
\bibitem{Buchdahl60} {Buchdahl~H.} {\em Am. J. Phys.} {\bf 28} (1960) 196. 
\bibitem{Buchdahl62} {Buchdahl~H.} {\em Z. Phys.} {\bf 168} (1962) 316. 
\bibitem{BuchdahlGreve} {Buchdahl~H. and Greve~W.} {\em Z.
Phys.} {\bf 168} (1962) 386. 
\bibitem{Boyling72} {Boyling~J.} {\em Proc. Roy. Soc. London}{ \bf A 329} (1972) 35. 
\bibitem{Zelenik76} {Zeleznik~F.} {\em J. Math. Phys.} {\bf 17} (1976) 1576.
\bibitem{Lieb99} {Lieb~E. and Yngvason~J.} {\em Phys. Rep.} {\bf 310} (1999) 1.
\bibitem{Prigogine67} {Prigogine~I.} {\em Introduction to
Thermodynamics of Irreversible Processes} (John Wiley \& Sons, New York) (1967).
\bibitem{Reidlich68} {Redlich~O.} {\em Rev. Mod. Phys.} {\bf 40} (1968) 556.
\bibitem{Bunge99} {Bunge~M.} {\em Dictionary of Philosophy}
(Prometheus Books, New York) (1999).
\bibitem{Lavoisier1780} {Lavoisier~A. et de Laplace P.}
{\em M\'emoire sur la chaleur. Memoirs de l'Academie des sciences} (1780) 355. 
(Reprinted in {\em Les Ma\^\i{}tres de la pens\'ee
    scientifique} (Gauthier-Villars, Paris) (1920).
\bibitem{Bunge74} {Bunge~M.} {\em Sense and Reference}
(Reidel, Dordrecht) (1974). 
\bibitem{Bunge67} {Bunge~M.} {\em Foundations of Physics}
(Springer-Verlag, New York) (1967).
\bibitem{Bunge77} {Bunge~M.} {\em Treatise of Basic
Philosophy. Ontology I: 
    The Furniture of the World} (Reidel, Dordrecht) (1977).
\bibitem{Bunge79} {Bunge~M.} {\em Treatise of Basic
Philosophy. Ontology 
    II: A World of Systems} (Reidel, Dordrecht) (1979).
\bibitem{Callen85} {Callen~H.} {\em Thermodynamics and an
Introduction to Thermostatics}, 2nd.Edn. (John Wiley \& Sons, New York) (1985). 
\bibitem{Galgani68} {Galgani~L. and Scotti~A.} {\em Physica}
{\bf 40} (1968) 150. 
\bibitem{Galgani69} {Galgani~L. and Scotti~A.} {\em Physica} 
{\bf 42} (1969) 242. 
\bibitem{Chen99} {Chen~M.} {\em J. Math. Phys.} {\bf 40} (1999) 830.

\end{thebibliography}
\end{document}